# Is seismicity operating at a critical point?

Shyam Nandan[1], Sumit Kumar Ram[2], Guy Ouillon[3] and Didier Sornette[2,4]

*[1]Swiss Seismological Service, ETH Zürich, Sonneggstrasse 5, 8092, Zürich, Switzerland*
*e-mail address: shyam4iiser@gmail.com*
*[2]Department of Management, Technology and Economics, ETH Zürich,*
*Scheuchzerstrasse 7, 8092, Zürich, Switzerland*
*[3]Lithophyse, 4 rue de l'Ancien Sénat, 06300 Nice, France*
*[4]Institute of Risk Analysis, Prediction and Management (Risks-X), Academy for Advanced Interdisciplinary Studies,*
*Southern University of Science and Technology (SUSTech), Shenzhen, China*

Seismicity and faulting within the Earth crust are characterized by many scaling laws that are usually interpreted as qualifying the existence of underlying physical mechanisms associated with some kind of criticality in the sense of phase transitions. Using an augmented Epidemic-Type Aftershock Sequence (ETAS) model that accounts for the spatial variability of the background rates $\mu(x,y)$, we present a direct quantitative test of criticality. We calibrate the model to the ANSS catalog of the entire globe, the region around California, and the Geonet catalog for the region around New Zealand using an extended Expectation-Maximization (EM) algorithm including the determination of $\mu(x,y)$. We demonstrate that the criticality reported in previous studies is spurious and can be attributed to a systematic upward bias in the calibration of the branching ratio of the ETAS model, when not accounting correctly for spatial variability. We validate the version of the ETAS model which possesses a space varying background rate $\mu(x,y)$ by performing pseudo prospective forecasting tests. The non-criticality of seismicity has major implications for the prediction of large events.

Earthquakes have fascinated and continue to capture the imagination and interest of physicists, as they express how the huge, unbridled forces of Nature can be organized according to remarkable regular statistical laws obeying power-law statistics. The late Per Bak, one of the fathers of the concept of "self-organized criticality," was fond of exclaiming in his talks: "I love this law," when referring to (i) the Gutenberg-Richter (GR) distribution of earthquake seismic moments because it is valid over several decades larger than most known power laws in physical and social sciences. Several other scaling laws further characterize seismicity: (ii) the Omori law (the rate of aftershocks decays as $\frac{1}{(t-t_m)^p}$ after a main shock that occurred at a time $t_m$), (iii) a spatial Green function quantifying the power-law decay of the influence of the main shock as a function of the distance to its aftershocks, (iv) a power fertility law of the average number of aftershocks triggered as a function of the magnitude of the main shock, (v) power-law distributions of the lengths of the faults on which earthquakes occur, (vi) fractal, multifractal or hierarchical scaling of the set of earthquake epicenters as well as fault networks, and so on. For physicists, these laws suggest the existence of underlying physical mechanisms associated with some kind of criticality in the sense of phase transitions and field theory with zero mass. Indeed, many proposals in this spirit have been put forward to rationalize these power laws: self-organized criticality [1, 2] critical point behavior before large earthquakes [3-8], based on accelerating seismic release [9-12], combinations of the two [13-16].

The evidence for the criticality of the Earth's crust is thus generally inferred from the presence of scale invariance and power-law scaling. But it is well-known that many other mechanisms can be at the origin of power-law scaling without the need for invoking criticality [17, 18]. In this context, the class of Epidemic-Type Aftershock Sequence (ETAS) models offer a direct path to calibrate and quantify the distance to the criticality of earthquake catalogs. ETAS models combine the scaling laws (i)-(iv) to formulate a statistical framework that reproduces many of the statistical features observed in seismicity catalogs and is often used as the benchmark to assess the skills of competing earthquake forecasting models [19]. The ETAS models have an exact branching process representation [20], in which background earthquakes supposed to be driven by the forces of plate tectonics (immigrants in the language of epidemic branching processes) can trigger cohorts of earthquakes (first-generation "daughters"), themselves triggering second-generation events and so on [21]. The ETAS models exhibit a transition determined by the control parameter $n$, called the branching ratio, which is both the average number of triggered events of first-generation per immigrant and the fraction of all triggered events of any generation in the catalog [22]. The transition occurring at the critical value $n = 1$ separates the sub-critical regime ($n < 1$), where the sequence of events is stationary, from the super-critical regime ($n > 1$) for which the number of earthquakes explodes exponentially with time with a finite probability [21]. The transition at $n = 1$ is characterized by the standard signatures of criticality, such as diverging rate $\mu/(1 - n)$ of events where μ is the background rate, nonlinear "susceptibility" at $n = 1$ [23] and various power-law statistics [24-27]. A lot of attention has been devoted to determining the empirical value of n, with most studies suggesting that $n$ is very close to 1 [28, 29]. In contrast, some other studies averaging over broad tectonic areas find lower numbers in the range 0.35 to 0.65 [30]. If $n \cong 1$, a significant portion of the Earth's crust would then function close to or even exactly at criticality, so that previous earthquakes trigger most observed earthquakes. If confirmed, this has far-reaching consequences for modeling, predicting and managing seismic risks.

Within the representation offered by ETAS models, determining whether the Earth crust is at criticality or not is crucially dependent on the ability to partition the observed seismicity clustering in space and time between the background rates and the triggered rates resulting from previous earthquakes. For instance, a misclassification of earthquakes as background events due to incomplete catalogs biases the calibrated branching ratio n downward [31, 32]. On the other hand, failing to account for spatial and temporal variations of the background rates leads to upward biases for $n$, as the variability may be incorrectly attributed to triggering. This effect has been recently demonstrated in the time domain for financial time series [33, 34]. Here, we revisit this question of the criticality of the Earth's crust by offering an augmented ETAS model that improves on state of the art by accounting self-consistently for the possible spatial dependence of the key parameters.

Before presenting our method and results, we briefly review previous related studies [35-38] that have modeled the spatial dependence of the background rate $\mu(x, y)$ using a kernel density estimation. Generally, these studies use gaussian kernels with adaptive bandwidth. These kernel density estimates are obtained iteratively using the algorithm proposed by [35]. However, for all practical purposes, these studies fix the parameter on which the adaptive bandwidth of kernels depends, stating that the choice of this crucial parameter, which dictates the smoothing of background intensity in space, is relatively unimportant. Other studies [30,39] pre-delineate regions in which all the parameters are assumed to be uniform and the parameters are then individually or jointly inferred in each of these regions. In some other studies, such as [40], the spatial probability density function (PDF) of background earthquakes is pre-estimated and then fed into the ETAS inversion machinery. Furthermore, [41, 29, 42] infer the optimal space variation of all ETAS parameters jointly, while more recently, [43] have proposed an alternative approach to non-parametrically model the space variation of background rate using a Gaussian process prior.

Here, we extend the iterative algorithm of [35] to jointly estimate the ETAS parameters and the kernel parameters used for obtaining the spatially variable background rate using the Expectation-Maximization (EM) algorithm of [39]. This extended algorithm leads to an augmented ETAS model that accounts specifically for the space variation of $\mu(x, y)$. Using this model, we show that the value of the branching ratio and other triggering parameters depend crucially on how the background rate is modeled. We demonstrate, using synthetic tests, that ignoring the spatial variation of the background rate leads to estimated parameters that are highly biased. However, when the spatial variation of the background rate is accounted for, the calibrated parameters are close to their true values and unbiased. We also document this bias effect in the case of three real catalogs. We then perform rigorous pseudo-prospective experiments and show that the ETAS model with spatially varying background rates significantly overperforms the ETAS model with uniform background rates. To the best of our knowledge, such direct comparisons have never been made between these two models. This gives us confidence that the extended ETAS model is superior to its spatially invariant counterpart, and thus reveals genuine characteristics of seismicity.

**Self-consistent estimation of spatially variable background rate using an extended EM algorithm.** Our implementation of the ETAS model expresses the seismicity rate $\lambda(t, x, y|H_t)$, at any time t and location $(x, y)$ and conditional upon the history $H_t$ of seismicity up to t, as

$$\lambda(t, x, y|H_t) = \mu(x, y) + \sum_{i:t_i<t} g(t - t_i, x - x_i, y - y_i, m_i) \qquad (1)$$

$\lambda(t, x, y|H_t)$ receives contributions from the background intensity function μ and from the sum over all past earthquakes that can trigger future earthquakes according to the triggering function given by

$$g(t - t_i, x - x_i, y - y_i, m_i) = \frac{K \exp[a(m_i - M_c)]\{t - t_i + c\}^{-1-\omega} e^{-\frac{t-t_i}{\tau}} * T_{norm} * S_{norm}}{\{(x - x_i)^2 + (y - y_i)^2 + d \exp[\gamma(m_i - M_0)]\}^{1+\rho}} \qquad (2)$$

This expression combines the fertility law $P(m) = K \exp[a(m_i - M_c)]$ that quantifies the expected number of first-generation aftershocks ($\geq M_c$) triggered by an earthquake with magnitude m, the Omori-Utsu law $\{t - t_i + c\}^{-1-\omega} e^{-\frac{t-t_i}{\tau}}$ and the spatial Green function, leading to the set $\phi = \{\mu, K, a, c, \omega, \tau, d, \gamma, \rho\}$ of parameters that characterize the ETAS model. $T_{norm}$ and $S_{norm}$ are normalization constants for the time and space kernels, ensuring that they are proper probability density functions (PDF).

With this formulation, the branching ratio is

$$n = \int_{M_c}^{M_{max}} P(m) \times f(m)\, dm \qquad (3)$$

defined as the expected number of aftershocks of the first generation triggered by an earthquake, averaged over all magnitudes. The averaging over magnitude is thus performed using the GR distribution $f(m) = b\, ln(10) \frac{10^{-bm}}{10^{-bM_c} - 10^{-bM_{max}}}, \forall\, M_c \leq m \leq M_{max}$. Denoting $\alpha = a/\ln 10$, this yields

$$n = \begin{cases} \dfrac{Kb\left(1 - 10^{-(b-\alpha)(M_{max}-M_c)}\right)}{(b - \alpha)\left(1 - 10^{-b(M_{max}-M_c)}\right)}, \forall\, \alpha \neq b \\ \dfrac{Kb \ln(10)\ (M_{max} - M_c)}{\left(1 - 10^{-b(M_{max}-M_c)}\right)}, if\ \alpha = b \end{cases} \qquad (4)$$

We consider two variants of the ETAS model: $ETAS_{\mu}$, which features a spatially uniform background rate $\mu$, and $ETAS_{\mu(x,y)}$,

which possesses a space varying background rate $\mu(x,y)$. In $ETAS_{\mu(x,y)}$, $\mu(x,y)$ is informed by the spatial positions of previous earthquakes. The proposed parameterization given below in Eq. (5) should not be confused with the triggering part of the ETAS model given by the second term in the r.h.s. of Eq. (1), which also involves a summation over previous earthquakes. Here, the guiding idea is that observed earthquakes occur more frequently where the background intensity is larger because the background events are, by definition, the sources of all observed seismicity. Even if many earthquakes are triggered by previous earthquakes, their locations are related to that of their background sources [44, 22].

To estimate $\mu(x,y)$, we extend the EM algorithm proposed by [39] as described in details in Text S1 of the supplementary material (SM). In this algorithm, $\mu(x,y)$ is inferred by smoothing the location of background earthquakes, which are dynamically obtained during the inversion using a power-law kernel [45] with a bandwidth $D$ and exponent $Q$, which are estimated along with $\phi$.

**Dataset.** We apply the extended EM algorithm (Text S1 of SM) to catalogs obtained from two sources: the Advanced National Seismic System (ANSS) and GeoNet. The ANSS catalog is used for two study regions: the entire globe and the region around California. For the area around New Zealand, we use the GeoNet catalog. The location of ~600,000 earthquakes ($M \geq 3$) for the entire globe between 1975-2020, as reported in the ANSS catalog, and of ~1.2 million and ~600,000 earthquakes ($M \geq 1$) between 1975-2020 in the study regions surrounding California and New Zealand, as reported by the ANSS and Geonet catalogs respectively, are shown in Figures S1-3 of the SM. Text S2 in SM also presents the method to select the magnitude of completeness $M_c$ for each catalog.

**Parameter calibration.** The results of the calibration of the two ETAS models, $ETAS_{\mu}$ and $ETAS_{\mu(x,y)}$, on the global, Californian, and the New Zealand catalogs are presented in Table 1. When going from the $ETAS_{\mu}$ model to the $ETAS_{\mu(x,y)}$ model, the most remarkable changes are that (i) the overall background rate increases by nearly 29, 3, and 16 times for the three catalogs, respectively, while consequently (ii) the branching ratio $n$ is substantially smaller for the $ETAS_{\mu(x,y)}$ model compared to the $ETAS_{\mu}$ model. The other parameters also show substantial changes. More specifically, the branching ratio is remarkably close to 1 for the three catalogs when calibrated with the $ETAS_{\mu}$ model, which would lead to the erroneous conclusion that the Earth crust is critical, as discussed in the introduction. In contrast, the $ETAS_{\mu(x,y)}$ model gives n≅ 0.45, 0.79, and 0.61 for the global, Californian, and the New Zealand catalogs, respectively, clearly excluding criticality and qualifying the Earth crust in the sub-critical regime. The difference between the two spatial intensity of background earthquakes inferred from the calibrations of the two models is vividly illustrated in Figure 1(a and b) in the case of the global catalog (and Figures S4 and S5 in the SM for the Californian and New Zealand catalogs). We find that not only is the overall background rate is different between the two models, but also that the spatial patterns of the density of background earthquakes differ substantially between the two modeling choices, as one can observe the much more refined localization of the background rates along plate boundaries in $ETAS_{\mu(x,y)}$.

**Synthetic tests of the bias in branching ratio $n$ due to uniform background rate.** We now demonstrate that a realistic synthetic catalog with a relatively smaller branching ratio as found with the full $ETAS_{\mu(x,y)}$ model calibrated on the real catalogs yields a spuriously large branching ratio when calibrated with the $ETAS_{\mu}$ model, and recovers the correct value when calibrated with the full $ETAS_{\mu(x,y)}$ model. For this, using the full $ETAS_{\mu(x,y)}$ model, we simulate a 50-year long synthetic catalog [46] with earthquakes of magnitudes $M \geq 5$ for the entire globe using the parameters that are listed in Table 1 corresponding to $ETAS_{\mu(x,y)}$. Using the first ten years of the catalog as the auxiliary period and the remaining as the primary period [47], we calibrate the $ETAS_{\mu}$ model on this synthetic catalog. The obtained parameters are: $N_{bkg} = 22.27\ year^{-1}, \log_{10} K = -0.18, \quad a = 1.07, \quad \log_{10} d = 2.17, \rho = 0.73, \gamma = 0.78, \log_{10} c = -3.09, 1+\omega = 0.80, \log_{10}\tau = 3.9, n = 1.17$ (see Table 1). This should be compared with the true input parameters $N_{bkg} = 1{,}013.2\ year^{-1}$ for the background rate and $n = 0.45$ for the branching ratio for the global catalog (Table 1). The background rate inverted with the $ETAS_{\mu}$ model is too low, and the inferred branching ratio is too large, being remarkably close to the value inferred by inverting this model with uniform background rate $\mu$ on the real catalog, on nearly all the parameters. For instance, the Omori exponents inferred for the synthetic catalog and the real catalog are 0.80 and 0.79, respectively. This provides an excellent self-consistent test and further supports the validity of our hypothesis that the background rate $\mu(x,y)$ is strongly varying in space. This suggests that it is absolutely essential to account for the non-uniform background rate to obtain unbiased parameter estimates. These synthetic tests also show that the biases are predictable.

**Pseudo-prospective forecasting experiments as a validation step for $ETAS_{\mu(x,y)}$.** We now proceed to show that $ETAS_{\mu(x,y)}$ leads to operationally better forecasts of future seismic activity by setting up several 30 day long pseudo-prospective forecasting experiments at different spatial resolutions and magnitude thresholds $M_t$ of the validation catalog. For details on these pseudo-prospective experiments, we refer the readers to Text S3 in the SM. For more general discussions on the importance of these experiments and their design, we refer the readers to [46, 47].

At a given spatial resolution and magnitude threshold ($M_t$), the log-likelihood score of a model during a given testing period is defined as $LL = \sum_{i=1}^{N} \ln P(n_i)$, where $P(n_i)$ is the probability of observing $n_i$ earthquakes in the $i^{th}$pixel during the testing period, and N is the total number of equal area pixels that tile the study region. In any pixel, the probability is constructed using the number of earthquakes observed in different simulated catalogs as described in Text S3 in the SM.

Once the likelihoods for two models are calculated, the information gain of one over the other is simply the difference of their log-likelihoods. Figure 2(a-b) shows the time-series of cumulative information gain (CIG) that the $ETAS_{\mu(x,y)}$ model obtains over the $ETAS_{\mu}$ model in the 368 experiments that we perform with the global catalog (see similar figures S6-8 of the SM for the time series of CIG at different spatial resolutions and $M_t$ for Global, Californian and New Zealand catalogs). At all spatial resolutions and magnitude thresholds, the $ETAS_{\mu(x,y)}$ substantially outperforms the $ETAS_\mu$ model for all the three study regions.

To quantify if this over-performance of the $ETAS_{\mu(x,y)}$ is statistically significant, we define the mean information gain as the average information that $ETAS_{\mu(x,y)}$ obtains over the 368 testing periods. We then test the null hypothesis that this mean information gain is significantly larger than 0 against the alternative that it is not, using the student's t-test. Figure 2(c-d) (and similar figures S9-11 for all spatial resolutions and $M_t$ for the Global, Californian, and New Zealand catalogs) confirm that the mean information gain of the $ETAS_{\mu(x,y)}$ over $ETAS_\mu$ is significantly larger than 0, thus rejecting the null hypothesis at significance levels much smaller than 0.01 in all cases.

**Discussion and Conclusions.** Our results demonstrate that, when the spatial variation of background rate is appropriately accounted for, we get a superior forecasting model and a branching ratio that is much smaller than 1.

At a fundamental level, the non-critical value of the branching ratio $n$ invites a reexamination of the physical picture we have of the brittle rupture process in the Earth's crust. Until now, ~~large~~ values of $n$ close to unity have been reported, suggesting that the loaded fault network is in a permanent critical state, compatible with the popular concept of self-organized criticality. The much lower value of $n$ that we estimate using more appropriate assumptions and a superior algorithm suggests that fault networks mainly evolve far from a critical point. This has major implications for the prediction of large events. Indeed, in the self-organized critical scenario, each event is no different from all others from a generating process viewpoint, making prediction impossible [48]. In contrast, if the fault network remains most of the time far from criticality, more sporadic singularities may appear via various possible mechanisms and announce upcoming catastrophic events. This suggests, for instance, the need for a reevaluation of the Accelerated Moment Release hypothesis [11], benefitting from the prior use of the $ETAS_{\mu(x,y)}$ fitting model to better eliminate the contribution of the uncorrelated part of seismicity to the total seismicity rate.

Finally, our finding that seismicity operates in a sub-critical regime should not be confused with the phenomenon of sub-critical rupture of a single fault [49], where the stress is below the elastic stress threshold and time-dependent plastic deformations and micro-cracking occur slowly at and around the crash tip and at asperities. The finite-time singularity ending sub-critical crack growth [50] is a one-rupture problem. In contrast, the criticality (and its absence) of a large collection of interacting earthquakes is a many-body problem.

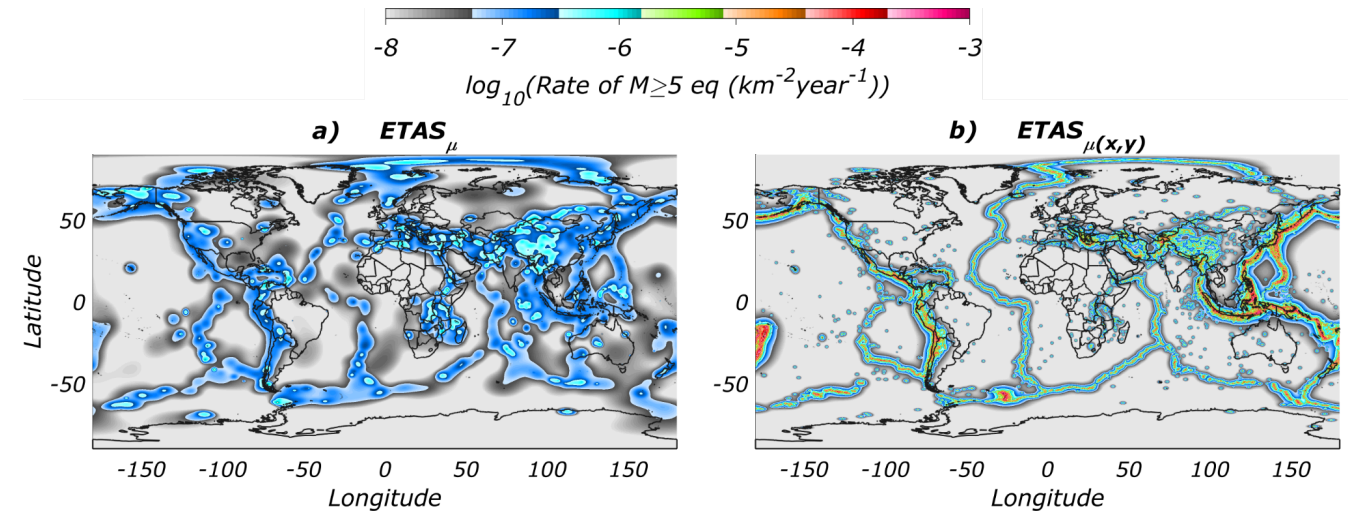

FIG 1: Spatial density of the earthquakes identified as background events by the $ETAS_\mu$ and $ETAS_{\mu(x,y)}$ models for the global catalog

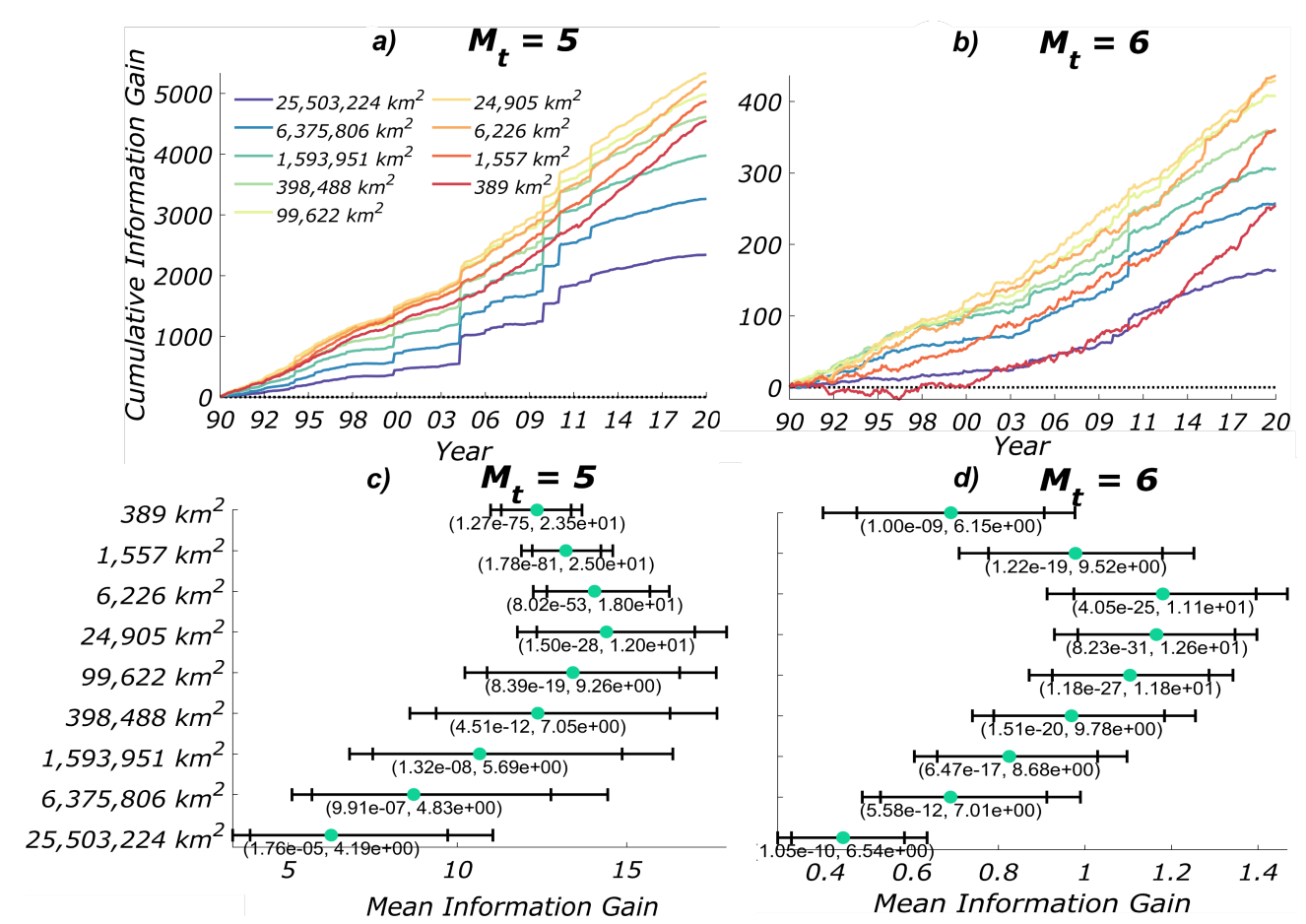

FIG 2: **(a,b)** Time series of cumulative information gain of $ETAS_{\mu(x,y)}$ over the $ETAS_\mu$ model in 368 pseudo prospective experiments with the global catalog, at nine spatial resolutions (different colors whose meaning is given in the inset) and two magnitude thresholds (different panels) of the testing catalog; **(c-d)** Mean information gain (MIG) of $ETAS_{\mu(x,y)}$ over the $ETAS_\mu$ model in 368 pseudo prospective experiments for the Global catalog, at nine spatial resolutions and two magnitude thresholds (different panels) of the testing catalog. The error bars indicate the 99 and 95 % confidence interval of the mean information gain (MIG). The two numbers indicate the p-value and t-statistic resulting from the student's t-test, in which we test the null hypothesis that the MIG is equal to 0 against the alternative that it is larger than 0. When the p-value is smaller than 0.05, the null hypothesis is rejected.

| Catalog | Model type | $N_{bkg}$ $(year^{-1})$ | $\log_{10} K$ | $a$ | $\log_{10} d$ $(km^2)$ | $\rho$ | $\gamma$ | $\log_{10} c$ (days) | $1+\omega$ | $\log_{10} \tau$ | Branching Ratio $n$ | $D$ (km) | $Q$ |
|---|---|---|---|---|---|---|---|---|---|---|---|---|---|
| Globe | $ETAS_\mu$ | 35.35 | -0.12 | 0.84 | 2.20 | 0.71 | 0.56 | -3.16 | 0.79 | 3.81 | 1.15 | NA | NA |
| | $ETAS_{\mu(x,y)}$ | 1,013.2 | -0.72 | 1.39 | 1.93 | 1.03 | 0.99 | -2.08 | 1.06 | 3.41 | 0.45 | 13.46 | 0.74 |
| California | $ETAS_\mu$ | 60.73 | -0.22 | 0.95 | -0.59 | 0.52 | 1.11 | -2.94 | 0.94 | 3.87 | 1.00 | NA | NA |
| | $ETAS_{\mu(x,y)}$ | 182.42 | -0.37 | 1.11 | -0.73 | 0.59 | 1.25 | -2.64 | 1.01 | 3.67 | 0.79 | 4.39 | 1.07 |
| New Zealand | $ETAS_\mu$ | 9.08 | -0.17 | 1.06 | 1.60 | 0.75 | 0.00 | -3.01 | 0.89 | 3.76 | 1.14 | NA | NA |
| | $ETAS_{\mu(x,y)}$ | 149.00 | -0.58 | 1.50 | 1.17 | 0.76 | 0.44 | -2.28 | 1.09 | 3.37 | 0.61 | 11.79 | 1.49 |

TABLE 1: ETAS parameters inverted for three catalogs (first column) using the two models $ETAS_{\mu}$ and $ETAS_{\mu(x,y)}$. The branching ratio $n$ is inferred using Eq.(4) with $M_{max} = 10, 8.5\ and\ 9$ for the Global, Californian, and New Zealand catalog, respectively. Parameters c and t are given in days. Nbkg is the total number of background events per unit time.